\documentclass{elsart}

\usepackage[]{graphicx}
\usepackage{amssymb}

\begin{document}

\begin{frontmatter}

\title{Magnetic phase diagram of the spin-1 two-dimensional $J_1$-$J_3$  Heisenberg model on a triangular lattice}

\author{P.~Rubin$^\dagger$\thanksref{mail}},
\author{A.~Sherman$^\dagger$},
\author{M.~Schreiber$^{\dagger\dagger}$}

\address{$^\dagger$Institute of Physics, University of Tartu,
Riia 142, 51014 Tartu, Estonia}
\address{$^{\dagger\dagger}$Institut f\"{u}r Physik, Technische
Universit\"{a}t, D-09107 Chemnitz, Germany}
\thanks[mail]{Corresponding author:
E-mail: rubin@fi.tartu.ee}

\begin{abstract}
The spin-1 Heisenberg model on a triangular lattice with the ferromagnetic nearest, $J_1=-(1-p)J,$ $J>0$, and antiferromagnetic third-nearest-neighbor, $J_3=pJ$,
exchange interactions
is studied in the  range of the parameter $0 \leqslant p \leqslant 1$. Mori's projection operator technique is used as a method, which retains the rotation symmetry of spin components and does not anticipate any magnetic ordering. For zero temperature several  phase transitions are observed. At $p\approx 0.2$ the ground state is transformed from the ferromagnetic spin structure into a disordered state, which in its turn is changed to an antiferromagnetic long-range ordered state with the incommensurate ordering vector ${\bf Q = Q^\prime} \approx \left(1.16, 0  \right)$ at  $p\approx 0.31$.  With the further growth of $p$ the ordering vector moves along the line ${\bf Q^\prime-Q_c}$ to the commensurate point ${\bf Q_c}=\left(\frac{2\pi}{3}, 0\right)$, which is reached at $p = 1$. The final state with an antiferromagnetic  long-range order can be conceived as four interpenetrating sublattices with the $120^\circ$ spin structure on each of them.
Obtained results are used for interpretation of the incommensurate magnetic ordering observed in NiGa$_2$S$_4$.

\end{abstract}


\end{frontmatter}

 This work was motivated by the recent synthesis of crystals  NiGa$_2$S$_4$ \cite{SNakatsuju} and  Ba$_3$NiSb$_2$O$_9$ \cite{Cheng,Serbyn}. These compounds demonstrate interesting magnetic properties, which are mainly determined by a two-dimensional triangular lattice of Ni$^{2+}$ ions with  spin $S=1$. Both systems are characterized by  a spin disorder at low temperature and strong antiferromagnetic interactions (the Curie-Weiss temperature is negative). In particular,
in the compound NiGa$_2$S$_4$
the experiment on neutron scattering  revealed the incommensurate short-range order \cite{SNakatsuju} with the in-plane correlation length equal to 6.9  lattice spacings at $T = 1.5 K$. The grain size of the sample was several orders of magnitude larger than the correlation length.
 To describe the observed order the $J_1$-$J_3$ classic Heisenberg model with the dominating third-nearest-neighbor (TNN) antiferromagnetic ($J_3$) and weak nearest-neighbor (NN) ferromagnetic interaction ($J_1$) was proposed  \cite{SNakatsuju}.
 The ratio $J_1/J_3$ can be fitted such that the incommensurate ordering vector $\bf Q_{cl}$  of the model coincides with the observed momentum
 $\bf Q_{exp}$ of the peak in the scattering intensity.
 The vector $\bf Q_{cl}$ is close to the commensurate ordering vector  in the case when the NN interaction vanishes. To estimate the values of the exchange constants in NiGa$_2$S$_4$  {\it ab initio} density functional  calculations
  were performed \cite{Mazin1}. It was shown that the second-nearest-neighbor exchange constant is negligibly small, while the TNN exchange constant $J_3$ is
 anomalously large. The NN exchange constant is smaller than $J_3$ and possibly ferromagnetic. Hence the  ratio of parameters of the classical $J_1$-$J_3$ model proposed in Ref.~\cite{SNakatsuju} agrees with  numerical estimations of Ref.~\cite{Mazin1}.
However,  due to quantum effects the classical model cannot give a comprehensive description for the system of $S = 1$ spins.
It is known that the magnetic phase diagram of the quantum Heisenberg model with competitive interactions is much richer. In particular, the phase diagram contains phases with the short range
order (SRO), which separates the long range ordered (LRO) phases \cite{AFB1, AFB2011, pla2010} (the classical phase diagram has only LRO states).
Earlier the  quantum $J_1$-$J_3$  model  was considered in  the Schwinger-Boson mean field approach \cite{Li1}, where a fitting
to experimental results \cite{SNakatsuju} was attempted.

In this article we consider the $S=1$ $J_1$-$J_3$ model on a triangular lattice with the ferromagnetic  NN  ($J_1=-(1-p)J<0, J>0$) and the antiferromagnetic
TNN ($J_3=pJ>0$) couplings.
In the following we use $J$ as the unit of energy.
The  frustration parameter $p$ changes from 0 to 1. We use Mori's projection operator technique \cite{Mori}, which retains the rotation symmetry of spin components and does not anticipate any magnetic ordering. The used method  does
not require  any approximate representations of the spin operators for
calculating the spin Green's function. In this approach, this  function is represented by a continued fraction. The elements of the fraction are calculated in a recursive procedure, which is similar to Lanczos' orthogonalization \cite{Sherman}. It was  found that the magnetic phase diagram of the model consists of the LRO ferromagnet at $|J_1| \gg J_3$, a disordered state, incommensurate LRO antiferromagnetic phases with ordering vectors  located  on the line $\bf Q^\prime - Q_c$ [(${\bf Q^\prime}\approx (1.16, 0),
{\bf Q_c}=(2\pi/3, 0)$] and a combination of  four $ 120^\circ$ spin structures on non-interacting enlarged triangular sublattices.
That spin texture corresponds to the case when the coupling between nearest neighbors vanishes ($J_1=0, p=1$). Thus in the range $0 \le p \le 1$ the system undergoes three phase transitions. We show that the considered model can describe the incommensurate short-range order observed in
NiGa$_2$S$_4$ at the value of the frustration parameter $p = 0.82$ and at low but finite temperature.

The Hamiltonian of the  model  reads
\begin{equation}\label{hamiltonian}
H=\frac{1}{2}\sum_{\bf nm}J_{\bf nm}\left(s^z_{\bf n}s^z_{\bf
m}+s^{+1}_{\bf n}s^{-1}_{\bf m}\right),
\end{equation}
where $s^z_{\bf n}$ and $s^\sigma_{\bf n}$ are the components of the
spin-1 operators ${\bf s_n}$, {\bf n} and {\bf m} label sites of the triangular lattice, $\sigma=\pm 1$. The spin-1 operators can be written as $s^z_{\bf n}=\sum_{\sigma=\pm1} \sigma |{\bf n}, \sigma \rangle\langle{\bf n},\sigma|$ and $s^\sigma_{\bf n}=\sqrt{2} (|{\bf n}, 0\rangle\langle{\bf n},-\sigma| +  |{\bf n}, \sigma\rangle\langle{\bf n},0|)$, where $|{\bf n},  \pm 1\rangle$ and $|{\bf n},0\rangle$ are site states with different spin projections. As mentioned above, we take into account the NN and TNN interactions,  $J_{\bf nm}=J_1\sum_{\bf a}\delta_{\bf n,m+a} + J_3\sum_{\bf A}\delta_{\bf n,m+A} $
with the vectors {\bf a}  and {\bf A}=2{\bf a} connecting the NN and TNN  sites.
Hereafter we use the lattice spacing $ a=|{\bf a}| $ as the unit of length.

The retarded Green's function reads
\begin{equation}\label{green}
 D({\bf k}t)=-i\theta(t)\langle[s^z_{\bf k}(t),s^z_{\bf
-k}]\rangle,
\end{equation}
where $s^z_{\bf k}=N^{-1/2}\sum_{\bf n}e^{-i{\bf kn}}s^z_{\bf n}$, $N$ is the number of sites, $s^z_{\bf k}(t)=e^{iHt}s^z_{\bf
k}e^{-iHt}$ and the angular brackets denote the statistical averaging.

We exploit Mori's projection operator technique \cite{Mori,Sherman} for calculating the Fourier transform of Kubo's relaxation function,
$$(\!( s^z_{\bf k}|s^z_{\bf -k})\!)_\omega=\int_{-\infty}^\infty
dt e^{i\omega t}(\!( s^z_{\bf k}|s^z_{\bf -k})\!)_t,\quad (\!( s^z_{\bf k}|s^z_{\bf -k})\!)_t=\theta(t)\int_t^\infty dt'\langle[s^z_{\bf k}(t'), s^z_{\bf - k}]\rangle.$$
The Fourier transform of Green's function (\ref{green}) can be obtained from this relaxation function using the relation
\begin{equation}\label{gk}
D({\bf k \omega})=\omega (\!( s^z_{\bf k}|s^z_{\bf
-k})\!)_\omega -(s^z_{\bf k},s^z_{\bf -k}),
\end{equation}
where  $(A,B)=i\int_0^\infty dt\langle[A(t),B]\rangle$. In this approach,  $(\!( s^z_{\bf k}|s^z_{\bf -k})\!)_\omega$ is represented as the continued fraction
\begin{equation}\label{cfraction}
(\!( s^z_{\bf k}|s^z_{\bf -k})\!)_\omega=\frac{\displaystyle(s^z_{\bf
k},s^z_{\bf -k})}{\displaystyle \omega-E_0-\frac{\displaystyle
V_0}{\displaystyle\omega-E_1-\frac{\displaystyle V_1}{\ddots}}},
\end{equation}
where the elements $E_n$  and $V_n$  of the fraction are determined from the recursive procedure
\begin{eqnarray}
&&[A_n,H]=E_nA_n+A_{n+1}+V_{n-1}A_{n-1},\quad E_n=([A_n,H],
A_n^\dagger)\,(A_n,A_n^\dagger)^{-1},\nonumber\\[-0.5ex]
&&\label{lanczos}\\[-0.5ex]
&&V_{n-1}=(A_n,A_n^\dagger)\,(A_{n-1},
A_{n-1}^\dagger)^{-1},
\quad V_{-1}=0, \quad A_0=s^z_{\bf k},\quad n=0,1,2,\ldots\nonumber
\end{eqnarray}
The operators $A_i$ obtained in the course of these calculations form an orthogonal set, $(A_i,A^\dagger_j)\propto\delta_{ij}$.

Using procedure (\ref{lanczos}) we get
\begin{eqnarray*}
&&E_0=(i\dot{s}^z_{\bf k},s^z_{\bf -k})(s^z_{\bf k},s^z_{\bf
-k})^{-1}=\langle[s^z_{\bf k},s^z_{\bf -k}]\rangle(s^z_{\bf k},s^z_{\bf
-k})^{-1}=0,\quad A_1=i\dot{s}^z_{\bf k},\\
&&V_0=6 J \Big[ -(1-p) C_1 [ \gamma({\bf {\small  k}})-1 ] + p \, C_{2a} [ \gamma(2{\bf k})-1 ] \Big](s^z_{\bf k},s^z_{\bf -k})^{-1}, \\%\quad
&&E_1=(i^2\ddot{s}^z_{\bf k},-i\dot{s}^z_{\bf -k})(i\dot{s}^z_{\bf
k},-i\dot{s}^z_{\bf -k})^{-1}=0,
\end{eqnarray*}
where $\gamma({\bf  k})=\frac{1}{3}\cos(k_x)+\frac{2}{3}
\cos\left(\frac{k_x}{2}\right) \cos\left(\frac{k_y \sqrt3}{2}\right)$
in the orthogonal
system of coordinates, $C_1 = \langle  s_{\bf n}^{+1} s_{\bf n +
a}^{-1} \rangle$ and $C_{2a} = \langle  s_{\bf n}^{+1} s_{\bf n +
A}^{-1} \rangle$  are the spin correlations  on the NN and TNN sites, respectively. At this point we interrupt the continued fraction and calculate $(s^z_{\bf k},s^z_{\bf -k})$. In thus taken approximation $V_1 \propto (A_2 ,A_2^\dagger) = 0$. From this equation we find
\begin{equation}\label{aii}
\langle[i^2\ddot{s}^z_{\bf k},-i\dot{s}^z_{\bf -k}]\rangle=36 J^2 \Big[ -(1-p) C_1 [\gamma({\bf k})-1] + p \, C_{2a}
[\gamma(2{\bf k})-1 ] \Big]^2 (s^z_{\bf k},s^z_{\bf -k})^{-1}.
\end{equation}
The quantity $i^2\ddot{s}^z_{\bf k}$ in the left-hand side of this
equation is a sum of terms of the type $s^z_{\bf l}s^{+1}_{\bf
n}s^{-1}_{\bf m}$. Following Refs. \cite{Kondo,Shimahara91}, we use the decoupling
$$s^z_{\bf l}s^{+1}_{\bf
n}s^{-1}_{\bf m}=\left[\alpha \langle s^{+1}_{\bf n}s^{-1}_{\bf
m}\rangle(1-\delta_{\bf nm})+\frac{4}{3}\delta_{\bf nm}\right]s^z_{\bf
l}$$
for the case $\bf l \ne m,n$. Here $\alpha$ is the vertex correction. In contrast to the case $S=\frac{1}{2}$ \cite{pla2005}, the terms with $\bf l = n$   or $\bf l = m$ do not cancel each other completely. For $S=1$ the residual terms read
$$
P_{\bf l}=\frac{1}{\sqrt{2}}\sum_{\bf m} J_{\bf lm}^2
\left(
|{\bf l},+1\rangle  \langle {\bf l},0| s_{\bf m}^- -
 |{\bf m},+1\rangle  \langle {\bf m},0| s_{\bf l}^- -
 s_{\bf l}^+ |{\bf m},0\rangle  \langle {\bf m},+1| +
 s_{\bf m}^+ |{\bf l},0\rangle  \langle {\bf l},+1|
\right).
$$
We neglect these terms in the following calculations taking into account that $\left\langle P_{\bf l} \right\rangle =0$.

Using this approximation for $i^2\ddot{s}^z_{\bf k}$, from
Eq.~(\ref{aii}) we find $(s^z_{\bf k},s^z_{\bf -k})$ and from
Eqs.~(\ref{gk}) and (\ref{cfraction}) we get
\begin{equation}\label{gfd}
D({\bf k}\omega)=\frac{6 J \Big[-(1-p) [\gamma({\bf k})-1]C_1
 +  p [\gamma(2{\bf
 k})-1] \, C_{2a} \Big]}{\omega^2-
 \omega^2_{\bf k}},
\end{equation}
where
\begin{eqnarray}\label{omega}
 \omega^2_{\bf k} &=& 36 J^2 \alpha \Bigg\{ \Bigg.
(1-p)^2 [\gamma({\bf k})-1 ]
\left[\frac{C_1}{6}+C_1\gamma({\bf k})-C_2 -\frac{2 (1-\alpha) }{9 \alpha} \right] \nonumber \\ \nonumber
&+&   \left. p^2 [\gamma(2{\bf k})-1]
\left[\frac{C_{2a}}{6}+C_{2a}\gamma(2{\bf k})  -C_2^\prime -\frac{2 (1-\alpha) }{9 \alpha} \right]   \right.  \nonumber \\
&-& p (1-p)  \Big[ [  1-\gamma({\bf k})  ] ( C^{\prime \prime} - \gamma(2{\bf k}) C_1  )    + [   1-\gamma(2{\bf k})] ( C^{\prime \prime} - \gamma({\bf k}) C_{2a})  \Big]  \Bigg. \Bigg\},
\end{eqnarray}
$ C_2 = \frac{1}{6}\left(\frac{4}{3} + 2\langle s_{\bf n}^{+1}s_{\bf n+ d}^{-1}\rangle +2 C_1 + C_{2a}\right) $,
$C^{\prime \prime} =  \frac{1}{6}\left( 2\langle s_{\bf n}^{+1}s_{\bf n+  r}^{-1}\rangle  +
2 \langle s_{\bf n}^{+1}s_{\bf n+ d}^{-1}\rangle + C_1 +\langle s_{\bf n}^{+1}s_{{\bf n}+ 3{\bf a}}^{-1}\rangle\right)$ and $C^{\prime }_2 = \frac{1}{6}\left(\frac{4}{3} + \langle s_{\bf n}^{+1}s_{{\bf n}+ 4{\bf a}}^{-1}\rangle + 2 C_{2 a}+ 2 \langle s_{\bf n}^{+1}s_{{\bf n}+ 2 {\bf d}}^{-1}\rangle\right), {\bf d}={\bf a}_1 + {\bf a}_2, {\bf r}= 2 {\bf a}_1 + {\bf a}_2$, where ${\bf a}_1 =(1,0)$ and ${\bf a}_2 =\left( \frac{1}{2}, \frac{\sqrt{3}}{2}\right)$ are the basis vectors of the triangular lattice. As follows from Eq.~(\ref{gfd}), the quantity $\omega_{\bf k}$ is the frequency of the spin excitations. From Eq.~(\ref{omega}) we see that this frequency  tends to
zero when $\bf k \rightarrow 0$.

To find the parameters $\alpha$, $C_1 $, $C_2$, ${C}'_2 $, $C_{2a}$ and $C^{\prime \prime}$  in Eqs.~(\ref{gfd})
and (\ref{omega}) we use the relation connecting the spin correlations with Green's function (\ref{gfd})
\begin{equation}
\label{eq1} \left\langle {s_{\rm {\bf n}}^{+1} s_{\rm {\bf m}}^{-1} }
\right\rangle =\frac{6J}{N}\sum\limits_{\rm
{\bf k}} {e^{i{\rm {\bf k}}\left( {{\rm {\bf n}}-{\rm {\bf m}}}
\right)}\frac{(p-1) [{\gamma }({\rm {\bf k}})-1 ] C_1 + p \,  [{\gamma }(2{\rm {\bf k}})-1 ] C_{2a}}{\omega _{\rm {\bf k}}
}\coth \left( {\frac{\omega _{\rm {\bf k}}}{2 T}} \right).}
\end{equation}
Five equations for $C_1 $, ${C}_2 $, ${C}'_2 $, $C_{2a}$, $C^{\prime \prime}$, which are derived from Eq.~(\ref{eq1}), and the equation
\begin{equation}
\langle s^{+1}_{\bf m} s^{-1}_{\bf m} \rangle =4/3,
\end{equation}
which follows from the constraint  $\langle  {\bf s^{2}_{\bf m} }\rangle =2$, form the closed set  for calculating all parameters
of Eqs.~(7) and (8) for a finite temperature.

First we  discuss briefly the ground state of the corresponding classical model \cite{Ryo}. The classical ground state
is the combination of the spiral configurations
\begin{equation}
{\bf S_{\bf n}}={\bf u} \cos({\bf Q}_{\bf cl}  \ {{\bf n}})+{\bf v} \sin({\bf Q}_{\bf cl} \  {{\bf n}}),
\end{equation}
where $\bf u$ and $\bf v$ are the arbitrary orthogonal unit vectors. The ordering vector ${\bf Q}_{\bf cl}$ is chosen from the condition
of minimal energy.
For small values of the frustration parameter, $0 \leq p \leq 0.2$,  the spin system is ferromagnetically ordered with   ${\bf Q}_{\bf cl}=0$
(see the inset in Fig.~1, in which the length of the ordering vector as a function of $p$  is shown).
For larger values of the frustration parameter ${\bf Q}_{\bf cl}$   becomes nonzero.
The energy minimum is achieved at six, in general case incommensurate, vectors
$ \left(\pm k,0\right), \left(\pm \frac{1}{2}k, \pm \frac{\sqrt{3}}{2} k\right)
 $  and  $    \left(\pm \frac{1}{2}k,  \mp \frac{\sqrt{3}}{2} k\right).$ One of these vectors coincides with the experimental ordering vector
 ${\bf Q}_{\bf exp} \approx (1.981, 0)$  at $p_{cl}=0.826$.
At $p=1$ the vectors ${\bf Q}_{\bf cl}$ become commensurate and their length is equal to $ \frac{2\pi}{3}$. Let us consider the spin
configuration with ${\bf Q}_{\bf cl}= \left(k,0\right) $. The dependencies of the scalar products
between the classical NN  $({\bf S_0 \cdot S_a})$ and TNN $({\bf S_0} \cdot {\bf S}_{2{\bf a}})$  spins on $p$ are shown in  Fig.~1.
These quantities are classical analogs of the spin-spin correlators $C_1 $ and $C_{2a}$.
In the pure ferromagnetic state ($|{\bf Q}_{\bf cl}|=0$) all spins are codirectional. At $p \gtrsim 0.2$ the angle between spins
$\bf S_0$  and ${\bf S}_{2{\bf a}}$ grows and spins become opposite in direction at $p \approx 0.45$. With further decrease of the ferromagnetic
coupling $J_1$ and rise of the antiferromagnetic interaction this angle diminishes. At $p=1$ when $J_1=0$ four triangular sublattices
with lattice spacing $2 a$ become
independent and the ground state is the combination of the $120^\circ$  structures on each sublattice. One can say that
 the ferromagnetic  interaction $J_1$ at $p \approx 0.45 $ promotes the increase of the antiferromagnetic correlations between  $\bf S_0$  and ${\bf S}_{2{\bf a}}$
 and suppresses the geometric frustration on the four sublattices.

Let us now turn to the quantum case. The numerical results were obtained by solving the system of equations (9),(10)
for the entire range of the frustration parameter $0 \le p \le 1$ and for the temperatures $T/J=0$ and 0.2 on  lattices up to $216\times216$ sites with
 periodic boundary conditions. For the solution of the mentioned set of six equations we used the Optimization toolbox of the Matlab package.
If the system has  LRO  at $T = 0 $ the summation
over the wave vector in (9) can be divided into the contribution yielded at the ordering vector $\bf k=Q$, which is proportional to the condensation part $C,$ and the
fluctuation contribution given by other wave vectors
\cite{Shimahara91}.
$C$ plays the role of the order parameter. In the case $|J_1| \gg J_3$ the ground state is ferromagnetic and the ordering vector  ${ \bf Q}=(0,0)$. An additional equation for  calculating $C = C_{F}$
 is the condition that  in the vicinity of the $\Gamma$ point $\omega_{\bf k}^2$ does not contain terms proportional to $k^2$. Solving this set of equations we found that the long range ferromagnetic order exists for $0< p \lesssim 0.2$.
 The dependence of  $C_{F}$ on $p$ is shown in Fig.~2. In the range  $0< p \lesssim 0.2$
 the ferromagnetic condensation part is practically constant, and it
  vanishes abruptly at
  $p \approx 0.2. $ For $T=0$ the spin-spin correlation functions $C_1$ and $C_{2a}$  also do not change in the ferromagnetic region (see Figs.~3 and 4) and it is evident that this behavior is analogous to the classical case (Fig.~1). $C_1$ and   $C_{2a}$ for $T/J=0.2$ are also shown in  Figs.~3 and 4, and their dependencies are  smoother. The evolution of the zero-temperature spin-excitation spectrum with $p$ is shown
    in Fig.~5. The sections of this dispersion along the $k_x$ axis are shown in Fig.~6 for different values of $p$.
   Notice that in Fig.~5(a), which corresponds to $p=0$, the dispersion is parabolic  near the $\Gamma$ point. Such a spectrum is typical for the ferromagnetic LRO. The flat region   near the $\Gamma$ point in Fig.~5(b) points to radical changes in the spectrum, which occur for
   $p > 0.2$. \\
\newline
The frequency of  magnetic excitations  for $T>0$ and $p > 0.2$ vanishes at the $\Gamma$  point, and it has a minimum at the
$\bf k$-vector, which is located on the line $ {\bf Q^\prime - Q_c}$ (hereafter we indicate only one of six symmetric directions in the Brillouin zone).
The frequency in the minimum is small but finite.
Such a minimum at $p = 0.52$ is shown in the inset of  Fig.~6 by the dashed line.
The minimum of the Fourier transform of the exchange coupling, which defines the classical ordering vector, is also situated on this
line. As mentioned above, this classical ordering vector is incommensurate  for $p \neq 1$. It can be supposed that the ground state of the quantum  model at  a large enough antiferromagnetic coupling $J_3$  also has the long-range antiferromagnetic order with an incommensurate ordering vector. To check this assumption the following procedure was performed.
Since a finite lattice with a discrete set of $\bf k$ points was used in our calculation, at first we determined the value of $p$,  at which the frequency minimum falls on some allowed incommensurate wave vector $\bf Q$  at  $T > 0$. Assuming that for $T=0$  the frequency vanishes at this wave vector, the condensation part $C = C_{AF}$  was calculated in the same manner as in the earlier works \cite{pla2010, Shimahara91}. The result is shown in Fig.~2: $C_{AF}$  is nonzero for $p \geq  0.31.$  The ordering vector corresponding to the boundary value $p \approx 0.31$  is ${\bf Q} = (1.16,0)$.  Thus, at $T=0$ in the range   $0.31 \leq p < 1$ the spin lattice has  incommensurate LRO.
Due to the interplay between
the  geometrical frustration and the exchange interactions the dependence of the condensation part on $p$ has a maximum at $p \approx 0.75$.
With  increasing   frustration parameter the ordering vector moves along the $k_x$ direction from ${\bf Q^\prime}$ to the commensurate point ${\bf Q_c}=\left(\frac{2\pi}{3}, 0 \right)$.  This wave vector is reached at $p = 1$,
as can be seen in  Fig.~6.  The spectra in Fig.~5(c) and Fig.5(d) correspond to the incommensurate order at $p=0.52$  and the commensurate order at $p=1$. The length of the incommensurate ordering vector ${\bf Q}$ at $p=0.82$ is equal to $1.978$, which is close to the experimentally observed
length $1.981$ of the
vector $\bf Q_{exp}$.
This frustration parameter $p$ is also close to the value $p_{\bf cl}$ obtained in the classical model.
  The LRO established at $p = 1$  can be conceived as four interpenetrating $120^\circ$ spin structures on  sublattices with twice
  as large lattice spacing.
 Spin orientations on the different sublattices are independent of one another. This is  seen in  Fig.~3 -- in this limit $C_1$, the correlation function between spins on different sublattices, vanishes. At the same time $C_{2a}$ tends to the value of spin-spin correlation  between nearest neighbors in the model with  nearest-neighbor  exchange on a triangular lattice \cite{pla2010} (see Fig.~4).

In addition to the ferromagnetic and the incommensurate antiferromagnetic phases the phase diagram contains
a region without any LRO.  As seen in Fig.~2, this SRO phase exists in the range  $0.2 \lesssim p \lesssim 0.31 $.
The LRO phases of the quantum and classical Heisenberg models are similar. The main difference is that in the quantum model these  phases are separated by the phase with  SRO. Thus,  at zero temperature with the variation of the frustration parameter the system undergoes the phase transitions
 from the ferromagnetic LRO  to the SRO, from the SRO to the  incommensurate antiferromagnetic LRO, which continuously transforms into the  commensurate LRO.

Above we found that the considered model has the state with the incommensurate ordering vector  ${\bf Q} = (1.978, 0)$, which is close to that observed in the crystal NiGa$_2$S$_4$. However, the state we obtained has the LRO at zero temperature, while the experimental result corresponds to the SRO state with the correlation length $\xi$  equal to 6.9 lattice spacings at $T = 1.5 K$ \cite{SNakatsuju}. In accord with the Mermin-Wagner theorem \cite{MW} the considered model has no LRO at $T > 0.$  We estimated   $\xi$ for $T \approx 2.8 K$  and found that it is an order of magnitude larger than the experimental one. In the crystal, the correlation length is influenced by the interlayer interaction, which is absent in the model. Besides, the difference between the experimental and calculated lengths may be connected with imperfections of the crystal structure and some other frustrating interactions. Nevertheless, obtained results demonstrate that the  $S=1$  $J_1$-$J_3$   Heisenberg model is able to describe key features of the experimental results in NiGa$_2$S$_4$. Notice also that quasilinear behavior near the minima of the spin-excitation dispersion explains the quadratic temperature dependence of the specific heat, observed in the crystal \cite{SNakatsuju}.
By our estimations, the range of the quadratic dependence extends to $T/J \approx 0.4$.
Besides, the shape of the calculated uniform susceptibility qualitatively reproduces the experimental data \cite{SNakatsuju}.

In summary, Mori's projection operator technique was used for
investigating the excitation spectrum and spin correlations of the two-dimensional $S=1$ $J_1$-$J_3$ Heisenberg  model on a triangular lattice,
 which takes into account the ferromagnetic
nearest-neighbor [$J_1=(-1+p) J < 0$] and the antiferromagnetic third-nearest-neighbor ($J_3 = p J > 0$) interactions. The character of the ground state depends on the frustration parameter $p$. In the range $0 < p \lesssim 0.2$, when the ferromagnetic coupling $J_1$  is larger than
$J_3$, the system is ferromagnetically ordered. At $p \approx 0.2$ the phase transition
into a state with  short-range order
occurs. The next transition takes place at $p \approx 0.31$; the system changes to the  long-range order with the incommensurate
ordering vectors ${\bf Q},$ which are located on the lines connecting the $\Gamma$
point and  corners of the hexagonal Brillouin zone. With the growth of $p$ these vectors move to the corners and reach them at $p=1$.
This commensurate state can be conceived as four interpenetrating sublattices with the $120^\circ$ spin structure on each of them. With $p\rightarrow 1$ the spin correlations between the sublattices subside.
The phases with the long-range order of the quantum Heisenberg model are similar to those in the analogous classical model. Additionally the quantum model has the phase with the short-range order, which separates the long-range ferromagnetic and incommensurate antiferromagnetic  phases.
The model is able to describe the state with the incommensurate short-range order observed in NiGa$_2$S$_4$.

\section*{Acknowledgements}

This work was partly supported by the
European Regional Development Fund
(Centre of Excellence "Mesosystems: Theory and Applications", TK114), the DAAD and the ESF Grant No.~9371.

\newpage
\begin{center}{\large Figures}\end{center}

 \begin{figure}[bhtp]
\begin{center}
\includegraphics[height=0.8\linewidth, angle=-90]{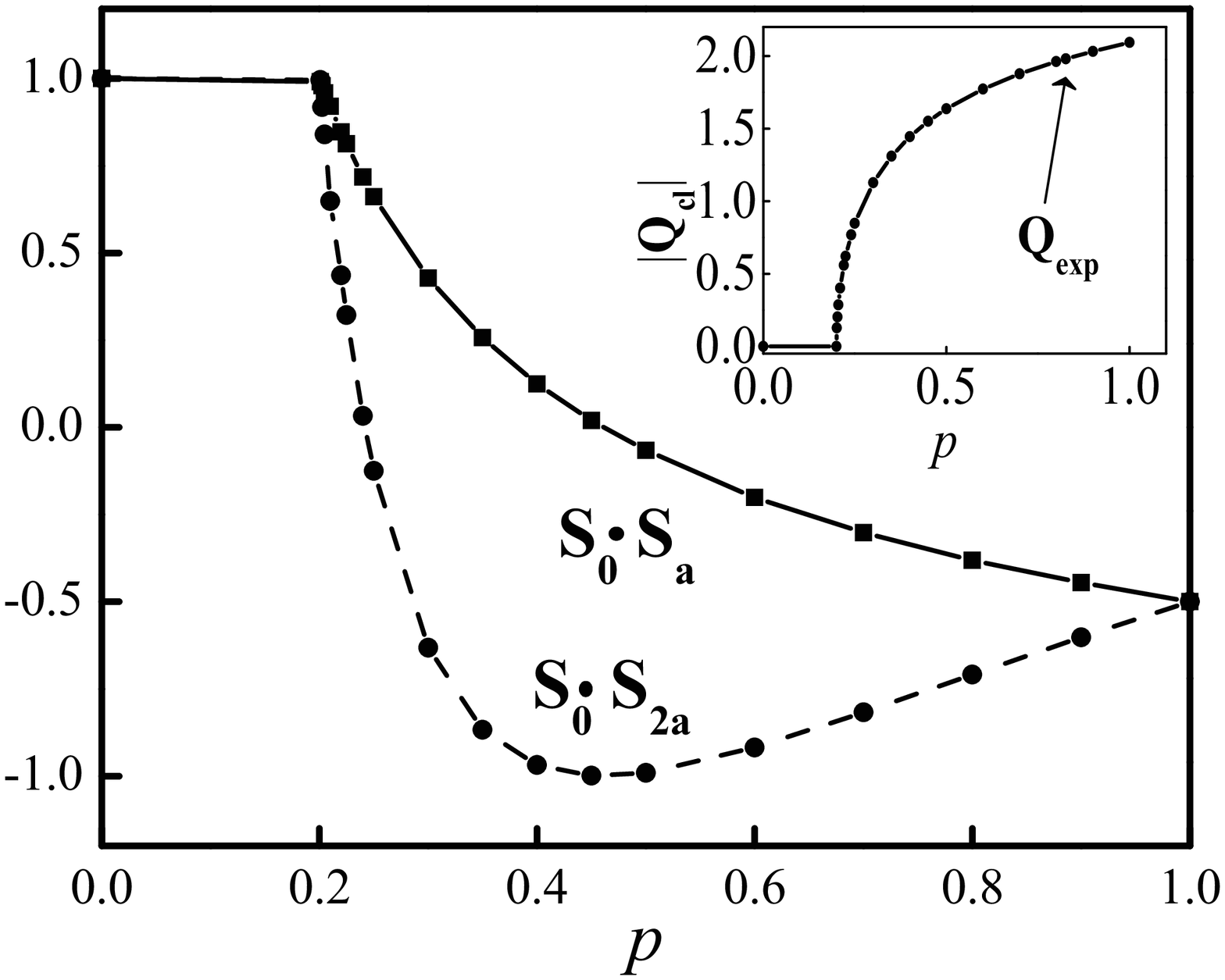}
\caption{ The dependencies of the scalar products
between the nearest  $({\bf S_0 \cdot S_a})$ and  third nearest neighbor $({\bf S_0 \cdot S_{2a}})$  classical spins on the frustration parameter $p$. The inset shows the dependence of the length of the ordering vector ${\bf Q}_{\bf cl}$ on $p$.}
\label{fig1}
\end{center}
\end{figure}

\newpage

 \begin{figure}[bhtp]
\begin{center}
\includegraphics[height=0.8\linewidth, angle=-90]{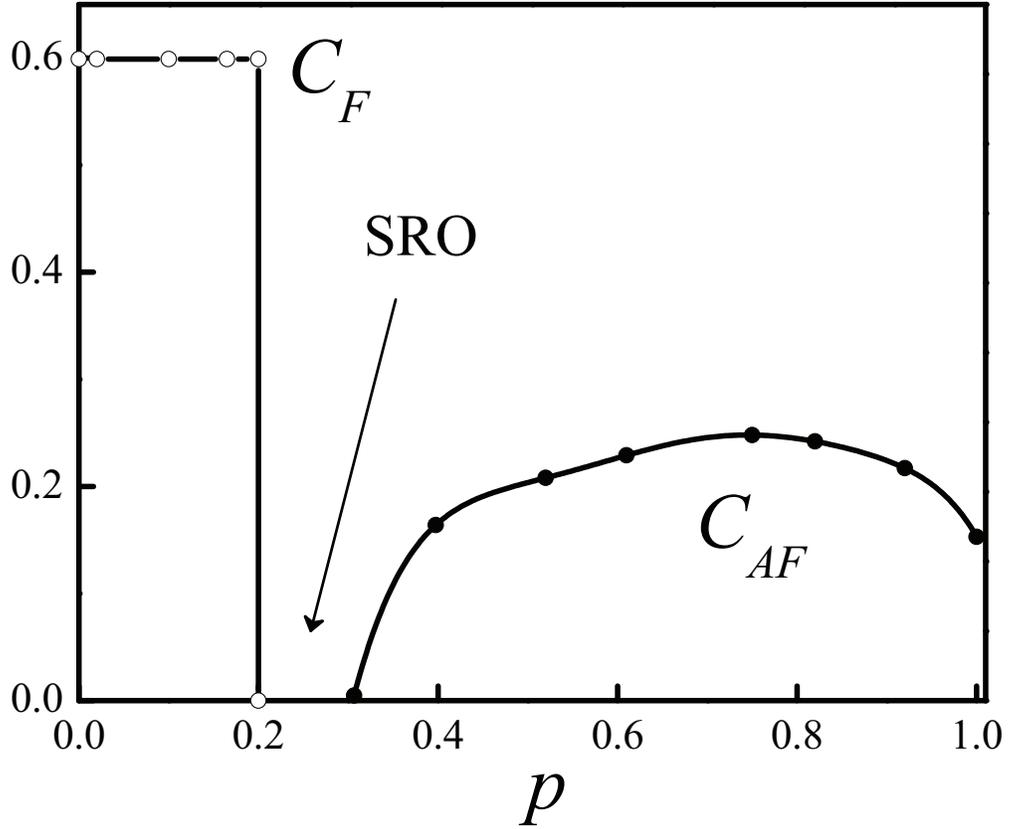}
\caption{ The dependencies of the  ferromagnetic condensation part $C_{F}$ (open circles) and the antiferromagnetic condensation part   $C_{AF}$ (filled circles) on the frustration parameter $p$.}
\label{fig1}
\end{center}
\end{figure}

\newpage

 \begin{figure}[bhtp]
\begin{center}
\includegraphics[height=0.8\linewidth, angle=-90]{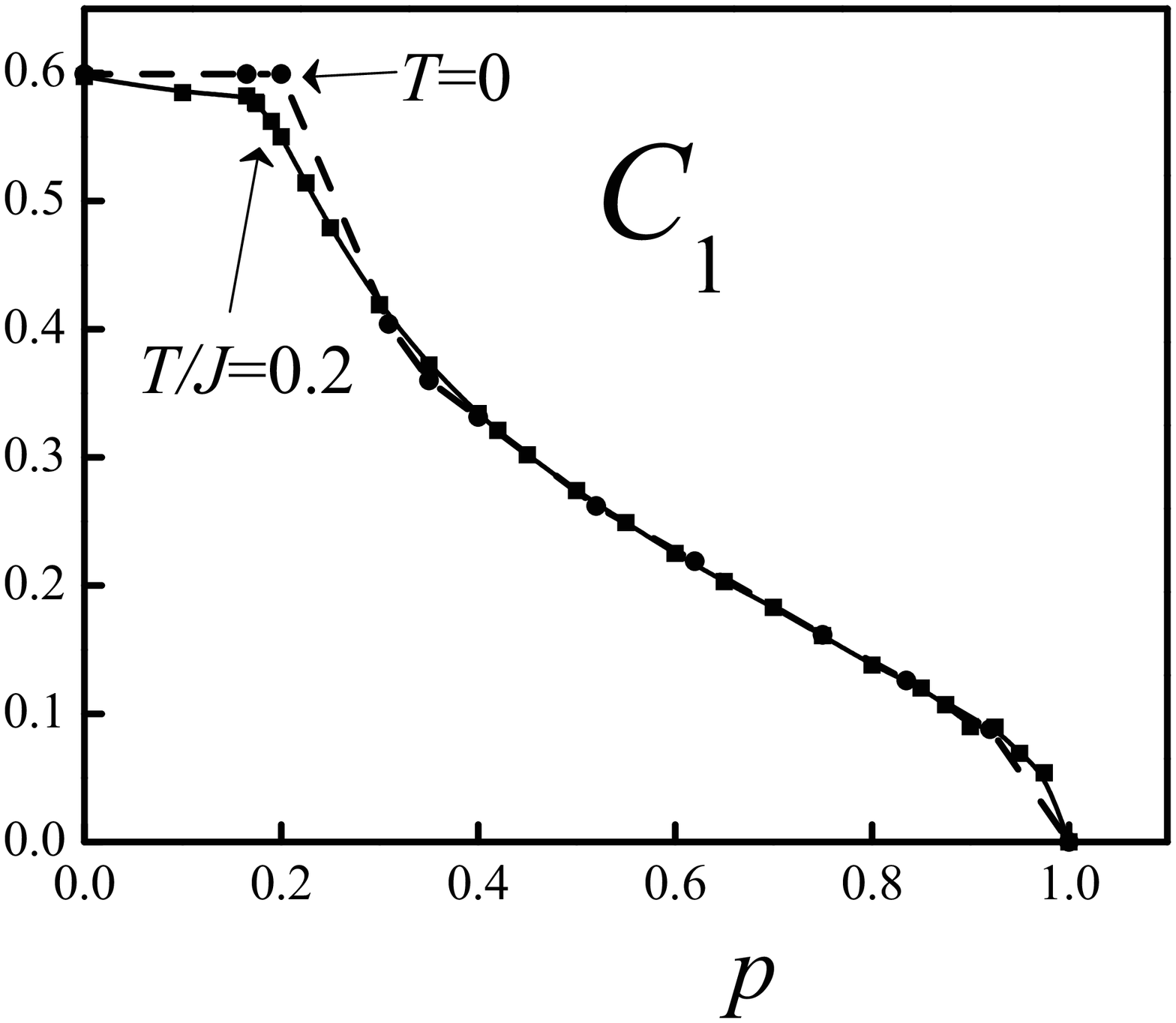}
\caption{ The dependence of the spin-spin correlation function between the nearest-neighbor spins  $C_1 = \langle  s_{\bf n}^{+1} s_{\bf n +
a}^{-1} \rangle$ on the frustration parameter $p$ at $T=0$ (circles, dashed line) and $T/J = 0.2$ (squares, solid line). }
\label{fig1}
\end{center}
\end{figure}

\newpage

 \begin{figure}[bhtp]
\begin{center}
\includegraphics[height=0.8\linewidth, angle=-90]{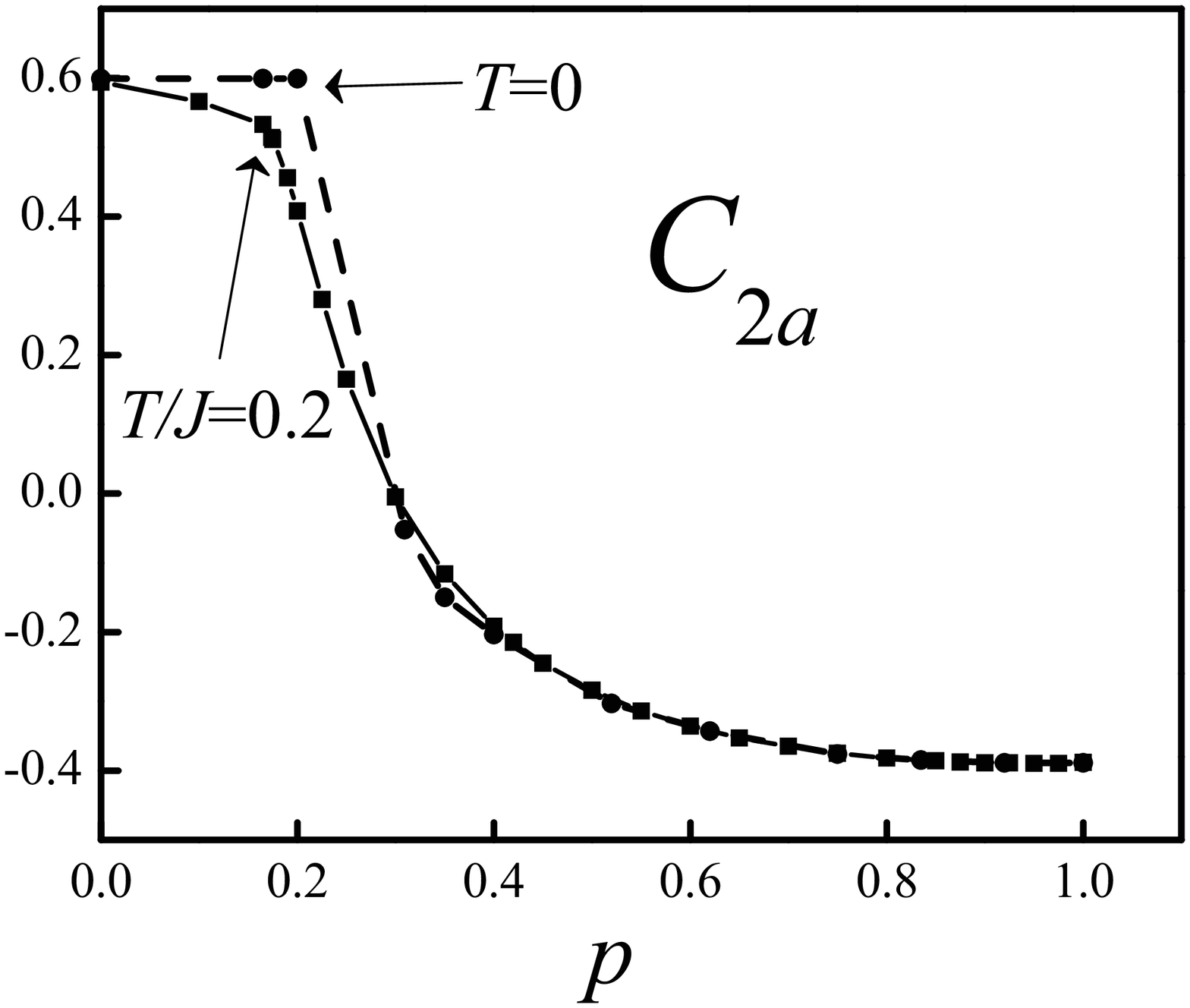}
\caption{ The dependence of the spin-spin correlation function between the third-nearest-neighbor spins $C_{2a} = \langle  s_{\bf n}^{+1} s_{\bf n +
A}^{-1} \rangle$
 on the frustration parameter $p$ at $T=0$ (circles, dashed line) and $T/J = 0.2$ (squares, solid line). }
\label{fig1}
\end{center}
\end{figure}

\newpage
\begin{center}{\large Figure 5 }\end{center}

{\large a)}

 \begin{figure}[bhtp]
\begin{center}
\includegraphics[height=0.8\linewidth, angle=-90]{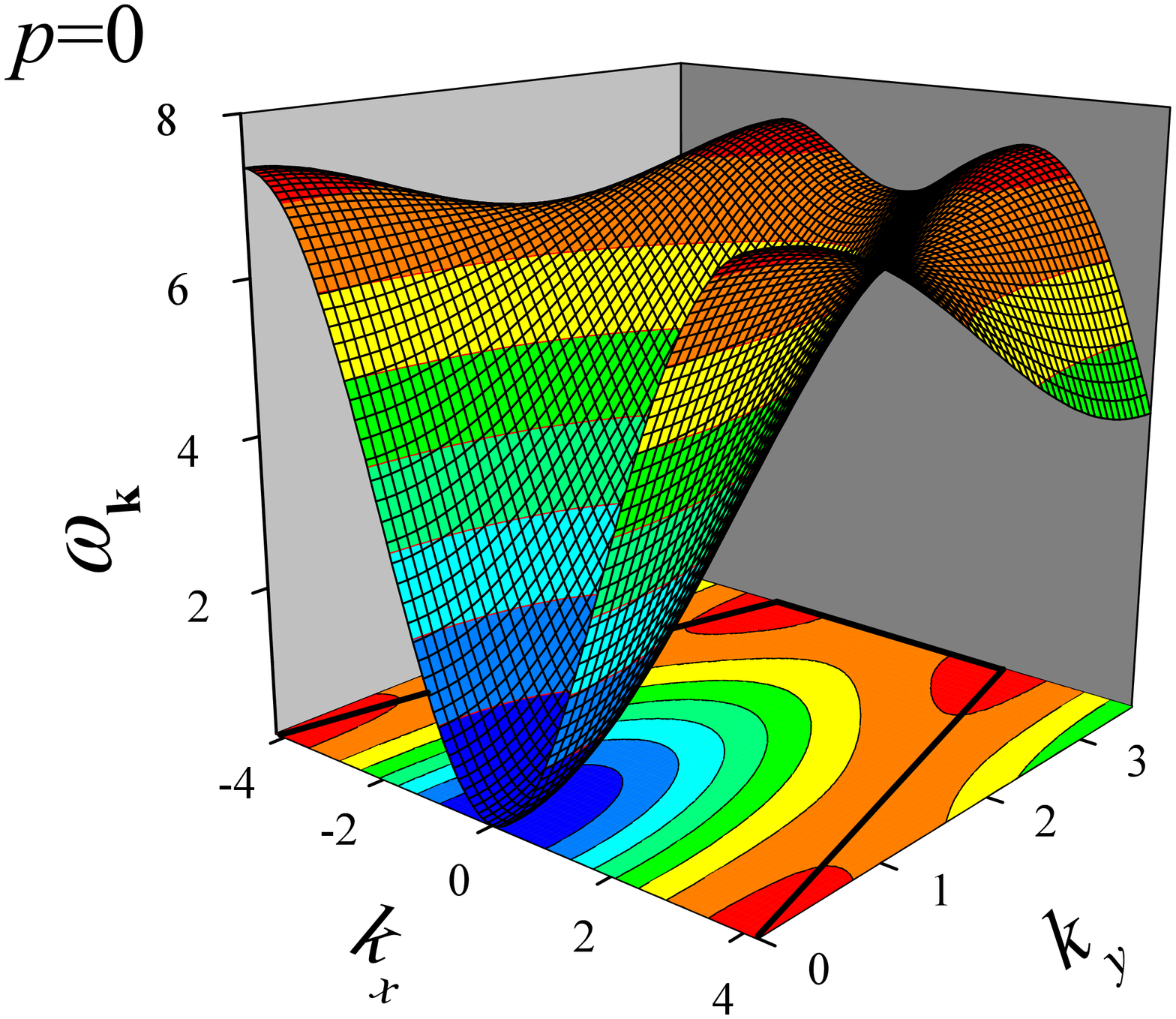}
\label{fig1}
\end{center}
\end{figure}

\newpage
\begin{center}{\large Figure 5 }\end{center}

{\large b)}

 \begin{figure}[bhtp]
\begin{center}
\includegraphics[height=0.8\linewidth, angle=-90]{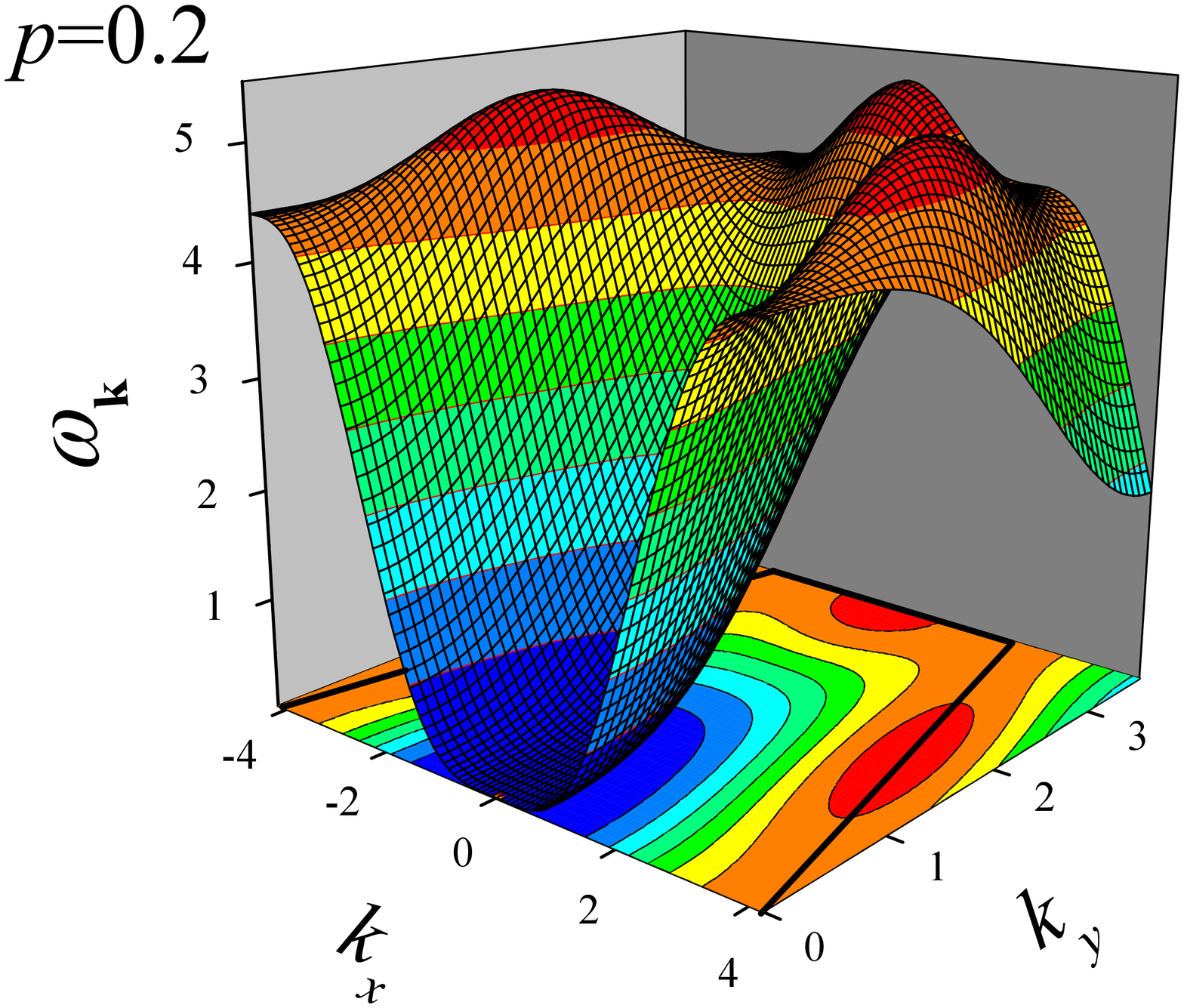}
\label{fig1}
\end{center}
\end{figure}

\newpage
\begin{center}{\large Figure 5 }\end{center}

{\large c)}

 \begin{figure}[bhtp]
\begin{center}
\includegraphics[height=0.8\linewidth, angle=-90]{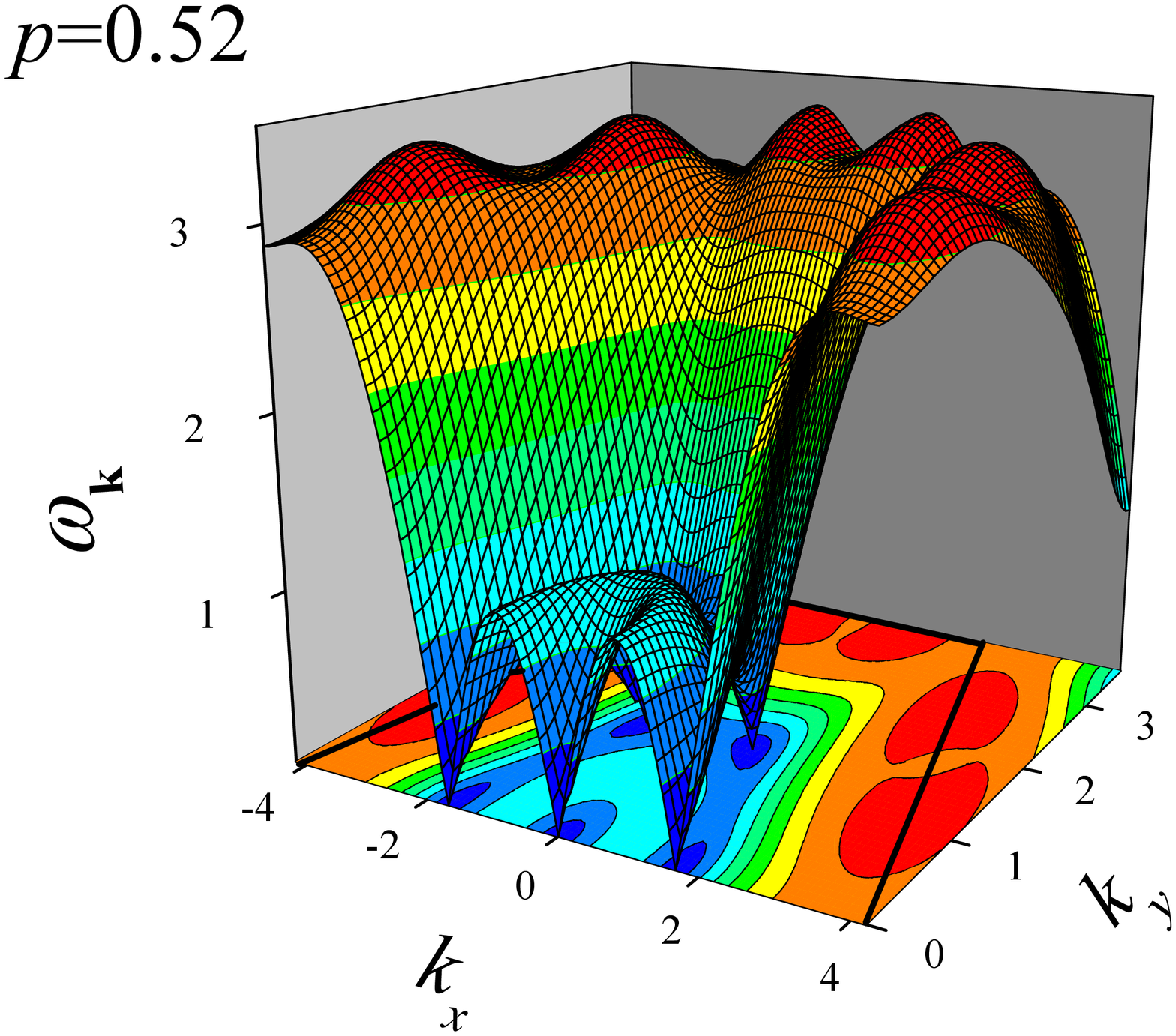}
\label{fig1}
\end{center}
\end{figure}

\newpage
\begin{center}{\large Figure 5 }\end{center}

{\large d)}

 \begin{figure}[bhtp]
\begin{center}
\includegraphics[height=0.8\linewidth, angle=-90]{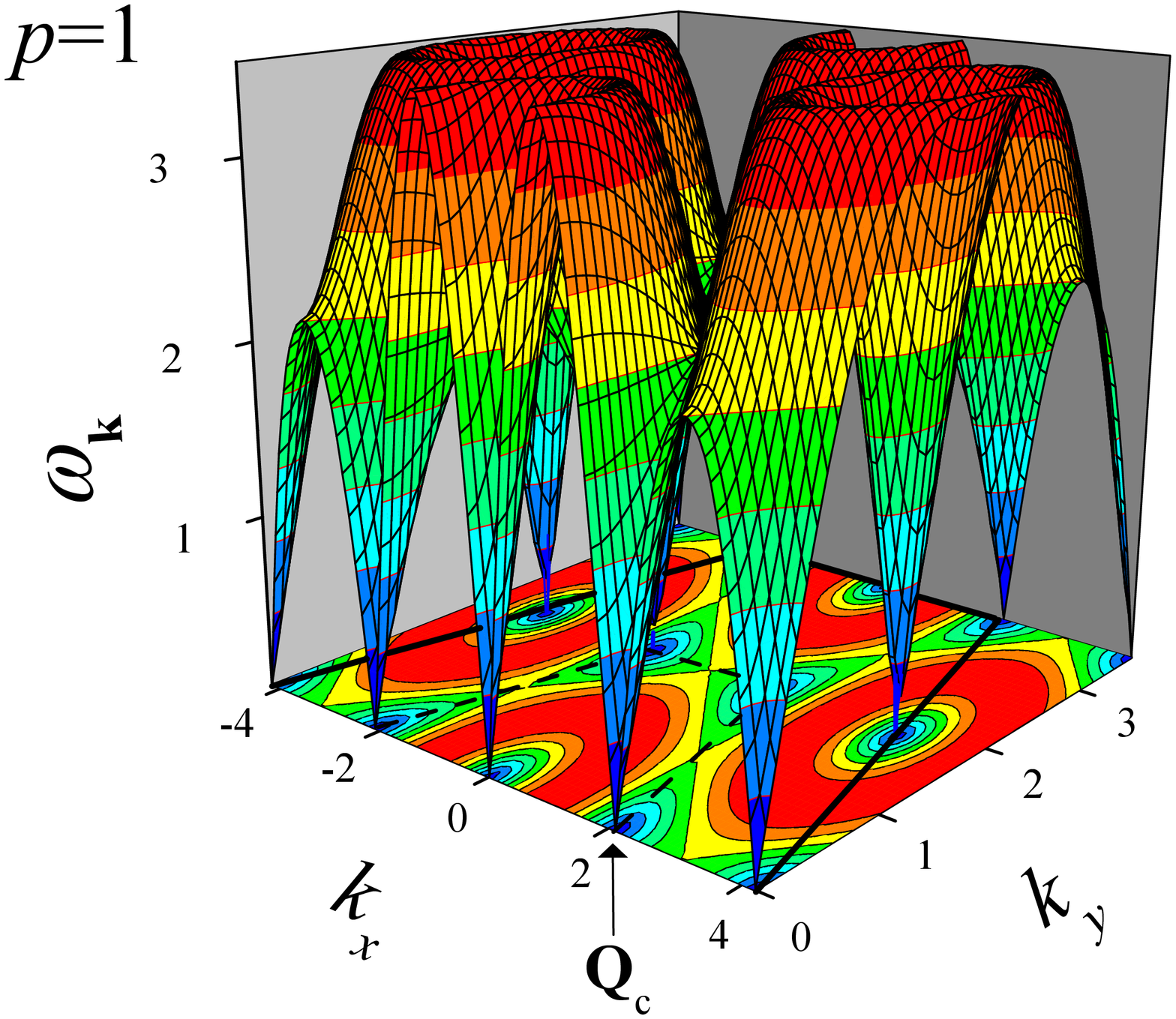}
\caption{The dispersion of spin excitations $\omega_{\bf k}$ for different values of the frustration parameter $p$ at $T=0$.
Half of the Brillouin zone is shown.
The thick solid line on the base plane is the
 border of the Brillouin zone for the triangular lattice. Panel (d) shows $\omega_{\bf k}$ for $p=1$. In that case the Hamiltonian
  splits into
  four terms, each of which describes
  the Heisenberg
  antiferromagnet on a
  triangular lattice with  twice as large period. The dashed line on the base plane shows one half of the Brillouin zone for this lattice.
  The ordering vector for $p=1$  is  ${\bf Q_c}=(2 \pi/3,0)$,
which lies at the corner of this small Brillouin zone.}
\label{fig1}
\end{center}
\end{figure}

\newpage

 \begin{figure}[bhtp]
\begin{center}
\includegraphics[height=0.8\linewidth, angle=-90]{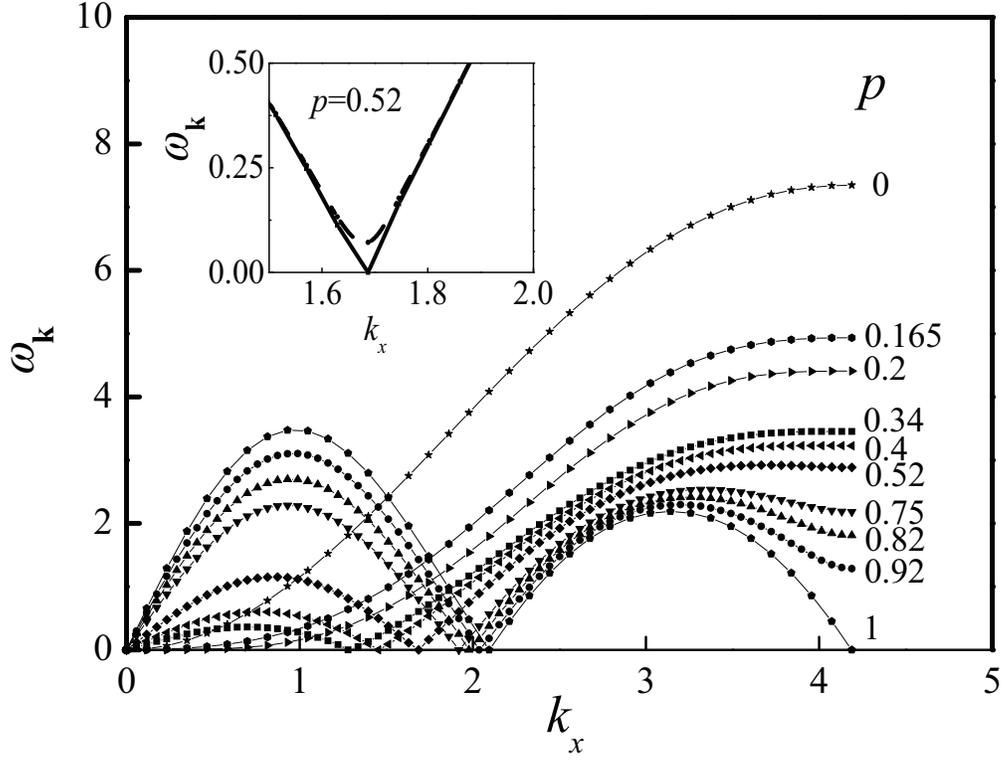}
\caption{ The dispersion of spin excitations $\omega_{\bf k}$ for different values of the frustration parameter $p$ at $T=0$ along the
$k_x$ axis. The inset shows  $\omega_{\bf k}$ at zero temperature (solid line) and  at $T/J=0.2$ (dashed line) for $p=0.52$. }
\label{fig1}
\end{center}
\end{figure}

\end{document}